\documentclass[aps,prl,twocolumn,groupedaddress,showpacs]{revtex4-1}
\usepackage{amssymb,amsmath,graphicx,epstopdf,hyperref,amsthm}

\begin{document}
\interfootnotelinepenalty=10000 

\title{Quantum information causality}

\author{Dami\'an Pital\'ua-Garc\'ia}
\affiliation{Centre for Quantum Information and Foundations, DAMTP, Centre for Mathematical Sciences,
University of Cambridge, Wilberforce Road, Cambridge, CB3 0WA, United Kingdom}


\begin{abstract}
How much information can a transmitted physical system fundamentally communicate? We introduce the principle of \emph{quantum information causality}, which states the maximum amount of quantum information that a quantum system can communicate as a function of its dimension, independently of any previously shared quantum physical resources. We present a new quantum information task, whose success probability is upper bounded by the new principle, and show that an optimal strategy to perform it combines the quantum teleportation and superdense coding protocols with a task that has classical inputs.
\end{abstract}

\maketitle
Quantum information science studies how information can fundamentally be encoded, processed and communicated via systems described by quantum physics \cite{NielsenandChuangbook}. Interesting features of information arise with this approach. The no-cloning theorem states that unknown quantum states cannot be copied perfectly \cite{WZ82,D82}. Unknown quantum states can be teleported \cite{teleportation}. Two classical bits can be encoded in one qubit via the superdense coding protocol \cite{sdc}. Fundamentally-secure cryptography can be achieved with quantum information protocols \cite{BB84,E91,BHK05}. Many of the quantum information protocols are possible due to quantum entanglement: two systems are entangled if their global quantum state cannot be expressed as a convex combination of individual states in a tensor product form. Another interesting property is quantum nonlocality, that is, measurement outcomes of separate systems can exhibit correlations that cannot be described by local classical models \cite{EPR35,Bell}.

Since the value of quantum correlations does not vary with the time difference of the measurements and the distance between the systems, one could think that they can be used to communicate arbitrarily-fast messages. However, quantum physics obeys the no-signaling principle. No-signaling says that a measurement outcome obtained by a party (Bob) does not provide him with any information about what measurement is performed by another party (Alice) at a distant location, despite any nonlocal correlations previously shared by them \cite{GRW80}.

If any information that Alice has is to be learned by Bob, no-signaling requires that a physical system sharing correlations with Alice's system must be transmitted to him. Thus, an interesting question to ask is: how much information can a physical system fundamentally communicate? In the scenario in which Alice has a classical random variable $X$, she encodes its value in a quantum state that she sends Bob and Bob applies a quantum measurement on the received state in order to obtain a classical random variable $Y$ as the output, the Holevo theorem \cite{K73} provides an upper bound on the classical mutual information between $X$ and $Y$. In the scenario in which Alice sends Bob $m$ classical bits, information causality states that the increase of the mutual information between Bob's and Alice's systems is upper bounded by $m$, independently of any no-signaling physical resources that Alice and Bob previously shared \cite{ic}. Information causality has important implications for the set of quantum correlations \cite{ic,ABPS09,CSS10,GWAN11,YCATS12}. For example, it implies the Cirel'son bound \cite{C80}, while the no-signaling principle does not \cite{PR94}.

Here we consider the scenario in which Bob receives a quantum system from Alice, who possibly shares quantum correlations with another party, Charlie, and ask the question: how much quantum information can Bob obtain about Alice's or Charlie's data? \footnote{A different question, investigated in Refs. \cite{CMMPPP12,SKB12} is how much entanglement can increase under local operations and quantum communication.} We introduce a new principle that we call \emph{quantum information causality}, which states that the maximum amount of quantum information that a quantum system can communicate is limited by its dimension, independently of any quantum physical resources previously shared by the communicating parties. Namely, the principle says that \emph{the increase of the quantum mutual information between Bob's and Charlie's systems, after a quantum system of $m$ qubits is transmitted from Alice to Bob, is upper bounded by $2m$}.

\begin{figure}
\includegraphics[scale=0.96]{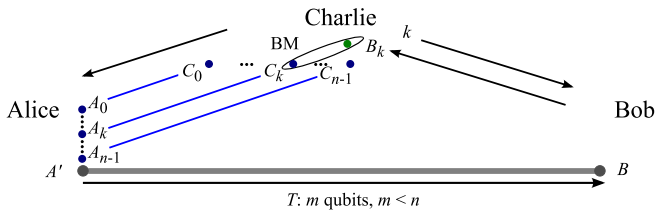}
 \caption{\label{fig1}(color online). The QIC game (version I).}
\end{figure}

In order to illustrate quantum information causality, we introduce a new quantum task that we call the \emph{quantum information causality (QIC) game} (see Fig.~\ref{fig1}).

\emph{The QIC game (version I)}. Initially, Alice and Bob may share an arbitrary entangled state. However, they do not share any correlations with Charlie. Let $A'$ and $B$ denote the quantum systems at Alice's and Bob's locations, respectively. Charlie prepares the qubits $A_j$ and $C_j$ in the singlet state $\lvert\Psi^-\rangle$, for $j = 0, 1,\ldots, n-1$. Charlie keeps the system $C \equiv C_0C_1\cdots C_{n-1}$ and sends Alice the system $A\equiv A_0A_1\cdots A_{n-1}$. Charlie generates a random integer $k\in\lbrace 0,1,\ldots,n-1\rbrace$ and gives it to Bob. Bob gives Charlie a qubit $B_k$, whose joint state with the qubit $C_k$, denoted as $\omega_k$, must be as close as possible to the singlet. Alice and Bob may play any strategy allowed by quantum physics as long as the following constraint is satisfied: their communication is limited to a single message from Alice to Bob only, encoded in a quantum system $T$ of $m < n$ qubits, with no extra classical communication allowed. Let $B'$ denote the joint system $BT$ after Bob's quantum operations. In general, the qubit $B_k$ is obtained by Bob from $B'$. Charlie applies a Bell measurement (BM) on the joint system $C_kB_k$. Alice and Bob win the game if Charlie obtains the outcome corresponding to the singlet. The success probability is
\begin{equation}
\label{eq:m2}
P\equiv\frac{1}{n}\sum_{k=0}^{n-1}\langle\Psi^-\rvert\omega_k\lvert\Psi^-\rangle.
\end{equation}

In version II of the QIC game, Charlie does not prepare singlets. Instead, Charlie prepares $n$ qubits in the pure states $\lbrace\lvert\psi_j\rangle\rbrace_{j=0}^{n-1}$ that he gives Alice. Bob outputs a qubit $B_k$ in the state $\rho_k$. Charlie measures $B_k$ in the orthonormal basis $\lbrace\lvert\psi_k\rangle,\lvert\psi_k^\bot\rangle\rbrace$. Alice and Bob win the game if Charlie's outcome corresponds to the state $\lvert\psi_k\rangle$. This version is equivalent to version I and its success probability $p$ satisfies: $p=(1+2P)/3$ (see details in the Supplemental Material). For convenience, in what follows we only refer to version I of the QIC game, unless otherwise stated.

Consider the following \emph{naive} strategy to play the QIC game. Alice simply sends Bob $m$ of the $n$ received qubits from Charlie without applying any operations on these. Alice and Bob previously agree on which qubits Alice would send Bob, for example, those with index $0 \leq j < m$. If Bob receives from Charlie a number $k < m$, he outputs the correct state; in this case, $\langle\Psi^-\rvert\omega_k\lvert\Psi^-\rangle=1$. However, if $m \leq k$, Bob does not have the correct state, hence, he can only give Charlie a fixed state, say $\lvert 0 \rangle$; in this case, $\langle\Psi^-\rvert\omega_k\lvert\Psi^-\rangle=1/4$. Thus, this strategy succeeds with probability $P_{\text{N}}=(1+3m/n)/4$, where the label N stands for \emph{naive}. There are other strategies that achieve success probabilities higher than $P_{\text{N}}$. However, it turns out that in general, $P < 1$, if $m < n$. We show that this follows from quantum information causality. 

The principle of \emph{quantum information causality} states an upper bound on the amount of quantum information that $m$ qubits can communicate:
\begin{equation}
\label{eq:m4}
\Delta I(C:B) \leq 2m,
\end{equation}
where $\Delta I(C:B) \equiv I(C:B')- I(C:B)$ is Bob's gain of quantum information about $C$, $I(C:B) \equiv S(C) + S(B)-S(CB)$ is the quantum mutual information \cite{NielsenandChuangbook} between $C$ and $B$, $S(C)$ is the von Neumann entropy \cite{NielsenandChuangbook} of $C$, etc., $B'$ denotes the joint system $BT$ after Bob's quantum operations. Since the quantum mutual information quantifies the total correlations between two quantum systems \cite{HV01,OZ02,GPW05}, we consider $\Delta I(C:B)$ to be a good measure for the communicated quantum information \footnote{Note that Refs.~\cite{HV01,OZ02,GPW05} propose measures for the purely classical and purely quantum parts of the correlations between two quantum systems, whose sum is equal to the quantum mutual information (see Ref. \cite{MBCPV12} for a review). We do not consider such a classification in our discussion.}.

The proof is very simple. By definition, $I(C:BT) = S(C) + S(BT)-S(CBT)$. Subadditivity \cite{LR68} states that $S(BT) \leq S(B) + S(T)$. The triangle inequality \cite{AL70}, $\lvert S(CB) - S(T)\rvert \leq S(CBT)$, implies that $-S(CBT) \leq S(T) - S(CB)$. Hence, we have that $I(C:BT) \leq 2S(T) + I(C:B)$. The data-processing inequality states that local operations cannot increase the quantum mutual information \cite{NielsenandChuangbook}. Thus, $I(C:B') \leq I(C:BT)$, which implies that $I(C:B') \leq 2S(T) + I(C:B)$. Therefore, we obtain that $\Delta I(C:B) \leq 2S(T)$. Finally, since $S(T) \leq \log_2 (\text{dim}T)$, the quantum information that $T$ can communicate is limited by its dimension. Therefore, if $T$ is a system of $m$ qubits, Eq.~(\ref{eq:m4}) follows because in this case $S(T) \leq m$. Achievability of equality in Eq.~(\ref{eq:m4}) requires that $T$ is maximally entangled with $C$ (see details in the Supplemental Material). It is easy to see that the naive strategy in the QIC game saturates this bound.

We notice that in the previous proof we did not require to mention Alice's system. This means that Eq.~(\ref{eq:m4}) is valid independently of how much entanglement Alice and Bob share. This also means that Eq.~(\ref{eq:m4}) is valid too if we consider that Alice and Charlie are actually the same party. Thus, quantum information causality shows: \emph{the maximum possible increase of the quantum mutual information between Charlie's and Bob's systems is only a function of the dimension of the system $T$ received by Bob, independently of whether it is Alice or Charlie who sends Bob the system $T$ and of how much entanglement Bob shares with them}.

If the transmitted system $T$ is classical, equality in Eq.~(\ref{eq:m4}) cannot be achieved. Information causality states that in this case, $\Delta I(C:B) \leq m$, where $C$ is a classical system, $B$ is a quantum system and $I(C:B)$ denotes their quantum mutual information \cite{ic}. In fact, this bound is valid even if both systems $C$ and $B$ are quantum (see details in the Supplemental Material). 

As stated above, quantum information causality follows from three properties of the von Neumann entropy: subadditivity, the data-processing and the triangle inequalities. The concept of entropy in mathematical frameworks for general probabilistic theories \cite{H01,B07,BBLW07} and its implication for information causality have been recently investigated \cite{SW10,BBCLSSWW10,AS11,DLR12}. Particularly, it has been shown that a physical condition on the measure of entropy implies subadditivity and the data-processing inequality, and hence that information causality follows from this condition \cite{AS11}. It would be interesting to investigate whether physically-sensible definitions of entropy for more general probabilistic theories satisfy the three mentioned properties, and hence a generalized version of quantum information causality. A different version of information causality in more general probabilistic theories has been considered in Ref.~\cite{MMAP12}.

Quantum information causality implies an upper bound on the success probability in the QIC game:
\begin{equation}
\label{eq:m5}
P \leq P',
\end{equation}
where we define $P'$ to be the maximum solution of the equation $h(P')+(1-P')\log_23=2(1-m/n)$ and $h(x)=-x\log_2x-(1-x)\log_2(1-x)$ denotes the binary entropy. The value of $P'$ is a strictly increasing function of the ratio $m/n$, achieving $P' = 1/4$ if $m = 0$ and $P' = 1$ if $m = n$. Therefore, we have that $P < 1$ if $m < n$. A plot with some values of $P'$ and the complete proof of Eq.~(\ref{eq:m5}) are given in the Supplemental Material. Below we present a sketch of the proof.

Firstly, we notice that for any strategy that Alice and Bob may play that achieves success probability $P$, there exists a covariant strategy achieving the same value of $P$ that Alice and Bob can perform. By covariance, we mean the following: in version II of the QIC game, if, when Alice's input qubit $A_k$ is in the state $\lvert\psi_k\rangle$, Bob's output qubit state is $\rho_k$, then, when $A_k$ is in the state $U\lvert\psi_k\rangle$, Bob's output state is $U\rho_kU^\dagger$, for any qubit state $\lvert\psi_k\rangle\in\mathbb{C}^2$ and unitary operation $U\in\text{SU(2)}$. Recall that $k$ is the number that Charlie gives Bob. Therefore, without loss of generality, we consider that a covariant strategy is implemented. This means that the Bloch sphere of the qubit $A_k$ is contracted uniformly and output in the qubit $B_k$. In version I, this means that the joint system $C_kB_k$ is transformed into the state
\begin{equation}
\label{eq:m7}
\omega_k=\lambda_k\Psi^-+\frac{1-\lambda_k}{3}\bigl(\Psi^++\Phi^++\Phi^-\bigr),
\end{equation}
where $1/4 \leq \lambda_k \leq 1$ and $\Psi^- $ denotes $\lvert\Psi^-\rangle\langle\Psi^-\rvert$, etc. That is, the depolarizing map \cite{NielsenandChuangbook} is applied to the qubit $A_k$, and output by Bob in the qubit $B_k$.

Then, we use the data-processing inequality and the fact that the qubits $C_j$ and $C_{j'}$ are in a product state for every $j \neq j'$ in order to show that $\sum_{k=0}^{n-1}I(C_k:B_k)\leq I(C:B')$. We notice that since Charlie's and Bob's systems are initially uncorrelated, Eq.~(\ref{eq:m4}) reduces to $I(C:B') \leq 2m$. Thus, we have that $\sum_{k=0}^{n-1}I(C_k:B_k)\leq 2m$. From this inequality and the concavity property of the von Neumann entropy, we obtain an upper bound on $\sum_{k=0}^{n-1}\lambda_k/n$, which from Eqs.~(\ref{eq:m2}) and~(\ref{eq:m7}) equals $P$.

Below we show that an optimal strategy to play the QIC game reduces to an optimal strategy to perform the following task.

\emph{The IC-2 game}. Alice is given random numbers $x_j \equiv (x_j^0,x_j^1)$, where $x_j^0,x_j^1\in\lbrace 0,1\rbrace$, for $j = 0, 1,\ldots, n- 1$. Bob is given a random value of $k = 0, 1,\ldots, n-1$. The game's goal is that Bob outputs $x_k$. Alice and Bob can perform any strategy allowed by quantum physics with the only condition that communication is limited to a single message of $2m < 2n$ bits from Alice to Bob. In particular, Alice and Bob may share an arbitrary entangled state. Let $y_k\equiv (y_k^0,y_k^1)$  be Bob's output, where $y_k^0,y_k^1\in\lbrace 0,1\rbrace$. We define the success probability as
\begin{equation}
\label{eq:m8}
Q\equiv \frac{1}{n}\sum_{k=0}^{n-1}P\left(y_k=x_k\right).
\end{equation}

We call this task the \emph{IC-2} game. The version we call the \emph{IC-1} game, in which the inputs and output are one bit values and Alice's message is of $m < n$ bits, was considered in the paper that introduced information causality \cite{ic}. The strategies to play the IC-1 game in which no entanglement is used were first considered by Wiesner in 1983 with the name of conjugate coding \cite{W83}. They were investigated further in 2002 with the name of random access codes (RACs) \cite{ANTV02}. The most general quantum strategy, in which Alice and Bob share an arbitrary entangled state, is called an entanglement-assisted random access code (EARAC) \cite{PZ10}. 

Let $Q_{\text{max}}$ be the maximum value of $Q$ over all possible strategies to play the IC-2 game. Below we show that $P\leq Q_{\text{max}}$.

Consider the following strategy to play the IC-2 game. Alice and Bob initially share a singlet state in the qubits $A_j$ and $C_j$, for $j = 0, 1,\ldots, n-1$. Alice has the system $A \equiv A_0A_1\cdots A_{n-1}$, while Bob has the system $C \equiv C_0C_1\cdots C_{n-1}$. Alice applies the unitary operation $\sigma_{x_j}$ on the qubit $A_j$, for every $j$, where $\sigma_{0,0}\equiv I$ is the identity operator acting on $\mathbb{C}^2$ and $\sigma_{0,1}\equiv \sigma_{1}$, $\sigma_{1,0}\equiv \sigma_{2}$, $\sigma_{1,1}\equiv \sigma_{3}$ are the Pauli matrices. Then, Alice and Bob play the QIC game, applying some operation on the input system $A$, which includes a message of $m$ qubits from Alice to Bob. However, instead of sending these $m$ qubits directly, Alice teleports \cite{teleportation} them to Bob. Thus, communication consists of $2m$ bits only, as required. At this stage, Bob does not apply any operations on the system $C$, which is consistent with the QIC game. As previously indicated, we can consider that in a general strategy in the QIC game the depolarizing map is applied to the qubit $A_k$. Therefore, Bob outputs the qubit $B_k$ in the joint state $\Omega_k=(I\otimes\sigma_{x_k})\omega_k(I\otimes\sigma_{x_k})$ with the qubit $C_k$, where $\omega_k$ is given by Eq.~(\ref{eq:m7}). Then, Bob measures $\Omega_k$ in the Bell basis. Bob learns the encoded value $x_k$ with probability $\lambda_k$. Thus, from Eq.~(\ref{eq:m8}) we have that $Q=\sum_{k=0}^{n-1}\lambda_k/n$, which equals $P$, as we can see from Eqs.~(\ref{eq:m2}) and~(\ref{eq:m7}). Since by definition $Q \leq Q_{\text{max}}$, we have that $P \leq Q_{\text{max}}$, as claimed.

Consider the following class of strategies to play the QIC game that combine quantum teleportation \cite{teleportation}, superdense coding \cite{sdc} (SDC) and the IC-2 game.

\emph{Teleportation strategies in the QIC game}. Alice and Bob share a singlet state in the qubits $A'_j$, at Alice's site, and $B_j$, at Bob's site, for $j = 0, 1,\ldots, n - 1$. Alice applies a Bell measurement on her qubits $A_jA'_j$ and obtains the two bit outcome $x_j \equiv (x_j^0,x_j^1)$. Thus, the state of the qubit $A_j$ is teleported to Bob's qubit $B_j$, up to the Pauli error $\sigma_{x_j}$. This means that the joint state of the system $C_jB_j$ transforms into one of the four Bell states, according to the value of $x_j$. Alice and Bob play the IC-2 game with Alice's and Bob's inputs being $x \equiv (x_0, x_1, \ldots, x_{n-1})$ and $k$, respectively. However, instead of sending Bob the $2m-$bits message directly, Alice encodes it in $m$ qubits via SDC.  Bob receives the $m$ qubits and decodes the correct $2m$-bits message, which he inputs to his part of the IC-2 game. Bob outputs the two bit number $y_k \equiv (y_k^0,y_k^1)$ and applies the Pauli correction operation $\sigma_{y_k}$ on the qubit $B_k$, which then he outputs and gives to Charlie. If $y_k = x_k$, the output state $\omega_k$ of the system $C_kB_k$ is the singlet; otherwise, we have that $\langle\Psi^-\rvert\omega_k\lvert\Psi^-\rangle=0$. Thus, from the definition of $P$, Eq.~(\ref{eq:m2}), we see that $P = Q$, where $Q$ is given by Eq.~(\ref{eq:m8}).

Therefore, since $P\leq Q_{\text{max}}$, we see that an optimal strategy in the QIC game is a teleportation strategy in which the IC-2 game is played achieving the maximum success probability $Q = Q_{\text{max}}$. We have obtained an upper bound on $Q$ for a particular class of strategies in the case $m = 1$ (see Supplemental Material).

The best strategy that we have found to play the QIC game in the case $m = 1$ is a teleportation strategy in which the IC-2 game is played with two equivalent and independent protocols in the IC-1 game. In both protocols Bob inputs the number $k$, while Alice inputs the bits $\lbrace x_j^0\rbrace_{j=0}^{n-1}$ in the first protocol and the bits $\lbrace x_j^1\rbrace_{j=0}^{n-1}$ in the second one. If Bob outputs the correct value of $x_k^0$ with probability $q$ in the first protocol, and similarly, he outputs the correct value of  $x_k^1$ with probability $q$ in the second protocol, for any $k$, then the success probability in the IC-2 game is $Q = q^2$. The maximum value of $q$ that has been shown \cite{AS11,PZ10} is $q = (1 + n^{-1/2})/2$. Explicit strategies to achieve this value are given by EARACs in the case in which $n = 2^r3^l$ and $r, l$ are nonnegative integers \cite{PZ10}. With this value of $Q$ we achieve a success probability in the QIC game of $P_{\text{T}}=\bigl(1+n^{-1/2}\bigr)^2/4$, where the label T stands for \emph{teleportation}.

Here we have introduced the quantum information causality principle as satisfaction of an upper bound on the quantum information that Bob can gain about Charlie's data as a function of the number of qubits $m$ that Alice (who shares correlations with Charlie) sends Bob, Eq.~(\ref{eq:m4}). We have presented a new quantum information task, the QIC game, whose success probability is limited by quantum information causality, Eq.~(\ref{eq:m5}). We have shown that an optimal strategy to play the QIC game combines the quantum teleportation and the quantum superdense coding protocols, with an optimal strategy to perform another task that has classical inputs, the IC-2 game. An optimal strategy in the IC-2 game remains as an interesting open problem.

\begin{acknowledgements}
I would like to thank Adrian Kent for much assistance with this work, and Nilanjana Datta, Sabri Al-Safi, Tony Short and Min-Hsiu Hsieh for helpful discussions. I acknowledge financial support from CONACYT M\'exico and partial support from Gobierno de Veracruz.
\end{acknowledgements}

\section{Supplemental Material}
\subsection{An equivalent version of the QIC game}

\emph{The QIC game (version II)}. This version is similar to version I, presented in the main text, with the following differences. Charlie does not prepare singlet states. Instead, Charlie prepares $n$ qubits in the pure states $\lbrace\lvert\psi_j\rangle\rbrace_{j=0}^{n-1}$, completely randomly. Charlie sends Alice the qubit $A_j$ in the quantum state $\lvert\psi_j\rangle$, for $j = 0, 1,\ldots, n-1$, and keeps a classical record of the states. We denote the global system that Alice receives from Charlie as $A \equiv A_0A_1\cdots A_{n-1}$. Bob gives Charlie a qubit $B_k$ in the state $\rho_k$, which must be as close as possible to $\lvert\psi_k\rangle$. Charlie measures the received state $\rho_k$ in the orthonormal basis $\lbrace\lvert\psi_k\rangle,\lvert\psi_k^\bot\rangle\rbrace$, where $\lvert\psi_k^\bot\rangle$ is the qubit state with Bloch vector antiparallel to that one of $\lvert\psi_k\rangle$. Alice and Bob win the game if Charlie's measurement outcome corresponds to the state $\lvert\psi_k\rangle$. The success probability is
\begin{equation}
\label{eq:m1}
p\equiv \int d\mu_0 \int d\mu_1 \cdots \int d\mu_{n-1} \biggl(\frac{1}{n}\sum_{k=0}^{n-1}\langle\psi_k\rvert\rho_k\lvert\psi_k\rangle\biggr),
\end{equation}
where $\int d\mu_j$ is the normalized integral over the Bloch sphere corresponding to the state $\lvert\psi_j\rangle$.

Now we show that both versions of the QIC game are equivalent and that their success probabilities satisfy the relation $p = (1 + 2P)/3$.  More precisely, we show that if Alice and Bob play a strategy in version I of the QIC game that achieves a success probability $P$, the same strategy applied to version II achieves a success probability $p$ that satisfies the relation $p = (1 + 2P)/3$, for any strategy that they may play, and vice versa.

We change to a more convenient notation, $\lvert\psi_k\rangle\equiv\lvert\uparrow_{\vec{r}_k}\rangle$, $\lvert\psi_k^\bot\rangle\equiv\lvert\downarrow_{\vec{r}_k}\rangle$, in order to make clear that $\lvert\psi_k\rangle$ and $\lvert\psi_k^\bot\rangle$ correspond to pure qubit states with Bloch vectors $\vec{r}_k$ and $-\vec{r}_k$, respectively. 

Version II of the QIC game is equivalent to the following. Charlie initially prepares the pair of qubits $A_j$ and $C_j$ in the singlet state $\lvert\Psi^-\rangle$, he gives Alice the qubit $A_j$ and keeps the qubit $C_j$, for $j = 0, 1, \ldots, n-1$. Charlie generates a random integer $k\in\lbrace 0, 1, \ldots, n-1\rbrace$ and gives it to Bob. Charlie measures the joint state $\omega_k$ of his qubit $C_k$ and the one received by Bob $B_k$ in the orthonormal basis $\mathcal{B}_{\vec{r}_k}\equiv\lbrace\lvert\uparrow_{\vec{r}_k}\rangle\lvert\uparrow_{\vec{r}_k}\rangle,\lvert\downarrow_{\vec{r}_k}\rangle\lvert\downarrow_{\vec{r}_k}\rangle,\lvert\uparrow_{\vec{r}_k}\rangle\lvert\downarrow_{\vec{r}_k}\rangle,\lvert\downarrow_{\vec{r}_k}\rangle\lvert\uparrow_{\vec{r}_k}\rangle\rbrace$ for some vector $\vec{r}_k$ that he chooses completely randomly from the Bloch sphere. Opposite outcomes correspond to success. Therefore, the success probability $p$ that Alice and Bob achieve in version II of the QIC game, given by Eq.~(\ref{eq:m1}), equals the following in this version:
\begin{align}
\label{eq:1}
p=&\int\!\! d\mu_0\!\int\!\! d\mu_1\cdots\!\int\!\! d\mu_{n-1}\!\biggl[\frac{1}{n}\sum_{k=0}^{n-1}\bigl(\langle\uparrow_{\vec{r}_k}\rvert\langle\downarrow_{\vec{r}_k}\rvert\omega_k\lvert\uparrow_{\vec{r}_k}\rangle\lvert\downarrow_{\vec{r}_k}\rangle \biggr. \biggr.\nonumber\\
&\qquad \biggl.  \bigl. +\langle\downarrow_{\vec{r}_k}\rvert\langle\uparrow_{\vec{r}_k}\rvert\omega_k\lvert\downarrow_{\vec{r}_k}\rangle\lvert\uparrow_{\vec{r}_k}\rangle\bigr)\biggr],
\end{align}
where $\int\! d\mu_j$ is the normalized integral over the Bloch sphere corresponding to the Bloch vector $\vec{r}_j$.

The Bell states defined in the basis $\mathcal{B}_{\vec{r}_k}$ are
\begin{eqnarray*}
 \lvert\Phi_{\vec{r}_k}^\pm\rangle&\equiv&\frac{1}{\sqrt{2}}\bigl(\lvert\uparrow_{\vec{r}_k}\rangle\lvert\uparrow_{\vec{r}_k}\rangle\pm\lvert\downarrow_{\vec{r}_k}\rangle\lvert\downarrow_{\vec{r}_k}\rangle\bigr),\\
 \lvert\Psi_{\vec{r}_k}^\pm\rangle&\equiv&\frac{1}{\sqrt{2}}\bigl(\lvert\uparrow_{\vec{r}_k}\rangle\lvert\downarrow_{\vec{r}_k}\rangle\pm\lvert\downarrow_{\vec{r}_k}\rangle\lvert\uparrow_{\vec{r}_k}\rangle\bigr).
\end{eqnarray*}
Consider that instead of measuring the state $\omega_k$ in the basis $\mathcal{B}_{\vec{r}_k}$, Charlie measures it in this Bell basis. Since the singlet state is the same in any basis, this corresponds to version I of the QIC game. Therefore, versions I and II of the QIC game are equivalent. Below we show that their success probabilities satisfy the claimed relation.

Using the Bell basis, we obtain from Eq.~(\ref{eq:1}) that
\begin{align}
\label{eq:2}
p=&\int\!\! d\mu_0\!\int\!\! d\mu_1\cdots\!\int\!\! d\mu_{n-1}\!\biggl[\frac{1}{n}\sum_{k=0}^{n-1}\bigl(\langle\Psi_{\vec{r}_k}^-\rvert\omega_k\lvert\Psi_{\vec{r}_k}^-\rangle \bigr. \biggr.\nonumber\\
&\qquad \biggl. \bigl. +\langle\Psi_{\vec{r}_k}^+\rvert\omega_k\lvert\Psi_{\vec{r}_k}^+\rangle\bigr)\biggr].
\end{align}
Since the singlet state $\lvert\Psi_{\vec{r}_k}^-\rangle$ is the same in any basis, by the definition of $P$ (Eq.~(\ref{eq:m2}) of the main text), we have that
\begin{equation}
\label{eq:3}
\int\!\! d\mu_0\!\int\!\! d\mu_1\cdots\!\int\!\! d\mu_{n-1}\frac{1}{n}\sum_{k=0}^{n-1}\langle\Psi_{\vec{r}_k}^-\rvert\omega_k\lvert\Psi_{\vec{r}_k}^-\rangle=P.
\end{equation}
On the other hand, we have that
\begin{eqnarray}
\label{eq:4}
\lefteqn{\int\!\! d\mu_0\!\int\!\! d\mu_1\cdots\!\int\!\! d\mu_{n-1}\langle\Psi_{\vec{r}_k}^+\rvert\omega_k\lvert\Psi_{\vec{r}_k}^+\rangle}\nonumber\\
 &=& \int\!\! d\mu_0\!\int\!\! d\mu_1\cdots\!\int\!\! d\mu_{n-1}\text{Tr}\bigl(\omega_k\lvert\Psi_{\vec{r}_k}^+\rangle\langle\Psi_{\vec{r}_k}^+\lvert\bigr)\nonumber\\
&=&\text{Tr}\biggl(\int\!\! d\mu_0\!\int\!\! d\mu_1\cdots\!\int\!\! d\mu_{n-1}\omega_k\lvert\Psi_{\vec{r}_k}^+\rangle\langle\Psi_{\vec{r}_k}^+\lvert\biggr)\nonumber\\
&=&\text{Tr}\biggl(\omega_k\int\!\! d\mu_0\!\int\!\! d\mu_1\cdots\!\int\!\! d\mu_{n-1}\lvert\Psi_{\vec{r}_k}^+\rangle\langle\Psi_{\vec{r}_k}^+\lvert\biggr)\nonumber\\
&=&\text{Tr}\biggl(\omega_k\int\!\! d\mu_k\lvert\Psi_{\vec{r}_k}^+\rangle\langle\Psi_{\vec{r}_k}^+\lvert\biggr),
\end{eqnarray}
where in the third line we have used the linearity of the trace; in the fourth line we have used the fact that $\omega_k$ does not depend on the Bloch vector $\vec{r}_k$ because Charlie chooses it completely randomly to define the measurement basis $\mathcal{B}_{\vec{r}_k}$, and can do so after Bob gives him the qubit $B_k$, and naturally does not depend on the Bloch vectors $\vec{r}_j$ with $j\ne k$ for the same reason; and in the last line we have used that the state $\lvert\Psi_{\vec{r}_k}^+\rangle$ is defined in terms of the Bloch vector $\vec{r}_k$, which is parameterized by $\mu_k$, and so is independent of the parameters $\mu_j$ with $j\neq k$.

It is easy to obtain that
\begin{equation}
\label{eq:5}
\int\!\! d\mu_k\lvert\Psi_{\vec{r}_k}^+\rangle\langle\Psi_{\vec{r}_k}^+\rvert=\frac{1}{3}\left(I-\lvert\Psi^-\rangle\langle\Psi^-\rvert\right),
\end{equation}
where $\lvert\Psi^-\rangle\equiv\bigl(\lvert 01\rangle-\lvert 10\rangle\bigr)/\sqrt{2}$ is the singlet state in the computational basis and $I$ is the identity operator acting on $\mathbb{C}^4$. From Eqs.~(\ref{eq:4}) and (\ref{eq:5}) and the definition of $P$ we have that
\begin{equation}
\label{eq:6}
\frac{1}{n}\sum_{k=0}^{n-1}\int\!\! d\mu_0\!\int\!\! d\mu_1\cdots\!\int\!\! d\mu_{n-1}\langle\Psi_{\vec{r}_k}^+\rvert\omega_k\lvert\Psi_{\vec{r}_k}^+\rangle=\frac{1}{3}-\frac{1}{3}P.
\end{equation}

Finally, we substitute Eqs.~(\ref{eq:3}) and~(\ref{eq:6}) into Eq.~(\ref{eq:2}) to obtain that $p = (1 + 2P)/3$, as claimed.

\subsection{Achievability of the quantum information causality bound}
We show that equality in Eq.~(\ref{eq:m4}) of the main text, $\Delta I(C:B)\leq 2m$, requires that the transmitted system $T$ is maximally entangled with Charlie's system $C$. 

Following the proof of Eq.~(\ref{eq:m4}) of the main text, we note that equality requires the following conditions to be satisfied. The transmitted system $T$ cannot be entangled with Bob's system $B$ in order to satisfy $S(BT)=S(B)+S(T)$. The system $T$ can only be entangled with the joint system $CB$ so that we have $-S(CBT)=S(T)-S(CB)$, as shown below. The state of the system $T$ has to be completely mixed so that its entropy is maximum: $S(T)=m$. This means that $T$ has to be maximally entangled with the system that purifies it. Together, these conditions imply that $T$ has to be maximally entangled with $C$. We also require that the quantum mutual information between $BT$ and $C$ does not decrease by Bob's operations: $I(C:B') = I(C:BT)$. 

Now we show that satisfaction of the equation $-S(CBT)=S(T)-S(CB)$ is achieved if and only if $T$ is entangled only with the joint system $CB$ \cite{NielsenandChuangbook}. Let $A$ be the quantum system that Charlie gives Alice, and hence is initially maximally entangled with $C$. Let any other physical system that Alice has to be denoted by $A'$. In particular, $A'$ can be entangled with Bob's system $B$, but not with Charlie's system $C$. Let $T$ be the system that Alice sends Bob. Since the systems $A'$ and $B$ are arbitrarily big, without loss of generality, we can consider that the global system $AA'CBT$ is in a pure state. Alice applies some quantum operation on the system $TAA'$, which in general can be represented by a unitary operation followed by a projective measurement. Thus, after Alice's operation, the global system $AA'CBT$ remains in a pure state. Due to the Schmidt decomposition of a bipartite pure state, we have that
\begin{eqnarray}
\label{eq:7}
S(CB)&&=S(TAA'),\nonumber\\
S(AA')&&=S(CBT).
\end{eqnarray}
We apply the subadditivity property to obtain
\begin{equation}
\label{eq:8}
S(TAA')\leq S(AA')+S(T),
\end{equation}
which from Eq.~(\ref{eq:7}) implies that
\begin{equation}
\label{eq:9}
S(CB)\leq S(CBT)+S(T).
\end{equation}
Equality in Eq.~(\ref{eq:9}) is achieved if and only if equality in Eq.~(\ref{eq:8}) is satisfied, which occurs if and only if $T$ is in a product state with $AA'$. Therefore, the relation $-S(CBT)=S(T)- S(CB)$ is satisfied if and only if $T$ is entangled only with the system $CB$, as claimed.

\subsection{The information causality bound}
If the transmitted system $T$ is classical, equality in Eq.~(\ref{eq:m4}) of the main text, $\Delta I(C:B)\leq2m$, can no longer be achieved. If $T$ represents a classical variable of $m$ bits then the smaller upper bound $\Delta I(C:B)\leq m$ is satisfied. The only difference in the proof of this bound compared to the one of $\Delta I(C:B)\leq2m$ is that if $T$ is classical then the bound $-S(CBT)\leq S(T)-S(CB)$ can no longer be saturated. In fact, in this case the smaller upper bound $-S(CBT)\leq-S(CB)$ is satisfied. A way to see this is that, if $T$ is a classical variable, the state of the joint system $CBT$ is a distribution over all possible values $x$ of $T$ and states of $CB$ for each $x$. Therefore, there exists a transformation $x\rightarrow(CB)_x$. From the data-processing inequality we have that $I(CB:T)\leq I(T:T)$. Hence, since $I(CB:T)=S(CB)+S(T)-S(CBT)$ and $I(T:T)=S(T)$, we obtain $S(CB)\leq S(CBT)$ \cite{ic}.

\subsection{Reduction of a general strategy in the QIC game to a covariant strategy}

For convenience, consider version II of the QIC game in which Charlie gives Alice $n$ pure qubits in the product state $\vec{\psi}\equiv\otimes_{j=0}^{n-1}\bigl(\lvert\psi_j\rangle\langle\psi_j\rvert\bigr)_{A_j}\in\mathcal{D}\Bigl(\bigl(\mathbb{C}^2\bigr)^{\otimes n}\Bigr)$, where we define $\mathcal{D}(\mathcal{H})$ to be the set of density operators acting on the Hilbert space $\mathcal{H}$. Let $\Gamma_k:\mathcal{D}\Bigl(\bigl(\mathbb{C}^2\bigr)^{\otimes n}\Bigr)\rightarrow\mathcal{D}\bigl(\mathbb{C}^2\bigr)$ be the map that Alice and Bob apply to the state $\vec{\psi}$, which outputs the state $\rho_k\equiv\Gamma_k\bigl(\vec{\psi}\!~\bigr)$ that Bob gives Charlie. Recall that $k$ is the number that Charlie gives Bob. After averaging over all possible input pure product states of qubits with index $j\neq k$, the output only depends on the state $\psi_k\equiv\lvert\psi_k\rangle\langle\psi_k\rvert$, which we identify with the map
\begin{multline}
\label{eq:10}
\bar{\Gamma}_k(\psi_k)\\
\equiv\!\int\!\!\! d\mu_0\!\!\int\!\!\! d\mu_1\!\cdots\!\!\int\!\!\! d\mu_{k-1}\!\!\int\!\!\! d\mu_{k+1}\!\!\int\!\!\! d\mu_{k+2}\!\cdots\!\!\int\!\!\! d\mu_{n-1}\Gamma_k\bigl(\vec{\psi}\!~\bigr),
\end{multline}
where $\int\!\! d\mu_j$ is the normalized integral over the Bloch sphere corresponding to the state $\lvert\psi_j\rangle$.

We define the map
\begin{equation}
\label{eq:11}
\bar{\Gamma}_k^{\text{cov}}(\phi)\equiv\int\!\! d\nu U_{\nu}^{\dagger}\bar{\Gamma}_k\bigl(U_{\nu}\phi U_{\nu}^{\dagger}\bigr)U_{\nu},
\end{equation} 
where $\phi\in\mathcal{D}\bigl(\mathbb{C}^2\bigr)$, $U_\nu\in\text{SU}(2)$ and $d\nu$ is the Haar measure on SU(2). It is easy to see that this map is covariant, that is, $\bar{\Gamma}_k^{\text{cov}}\bigl(U\phi U^\dagger\bigr)=U\bar{\Gamma}_k^{\text{cov}}(\phi)U^\dagger$, for all $\phi\in\mathcal{D}\bigl(\mathbb{C}^2\bigr)$ and $U\in\text{SU}(2)$.
 
In principle, for any map $\Gamma_k$ that Alice and Bob perform, they can implement the covariant map $\bar{\Gamma}_k^{\text{cov}}$  as follows. Alice and Bob initially share randomness. With uniform probability, they obtain the random number $\nu$ in the range $d\nu$ that corresponds to an, ideally,  infinitesimal region of the Haar measure on SU(2). This can be done, for example, if Alice and Bob share a maximally entangled state of arbitrarily big dimension and they both apply a local projective measurement in the Schmidt basis on their part of the state; their measurement outcome indicates the number $\nu$. Alice applies the unitary operation $U_\nu$ parameterized by the obtained number $\nu$ on each of her input qubit states $\lvert\psi_j\rangle$. Then, Alice and Bob apply the map $\Gamma_k$ to the input state $\otimes_{j=0}^{n-1}\bigl(U_\nu\lvert\psi_j\rangle\langle\psi_j\rvert U_\nu^\dagger\bigr)_{A_j}$. Finally, Bob applies the unitary $U_\nu^\dagger$ to his output qubit. From Eq.~(\ref{eq:10}) we obtain that, after averaging over all possible input pure qubits states with index distinct to $k$ and after Bob's final unitary operation $U_\nu^\dagger$, Bob's output state is $U_\nu^\dagger\bar{\Gamma}_k\bigl(U_\nu\psi_kU_\nu^\dagger\bigr)U_\nu$. Averaging over all shared random numbers $\nu$, we obtain $\bar{\Gamma}_k^\text{cov}(\psi_k)$, as defined by Eq.~(\ref{eq:11}). 

It is straightforward to see that the map $\bar{\Gamma}_k^\text{cov}$  satisfies
\begin{equation}
\int\!\! d\mu_k\langle\psi_k\rvert\bar{\Gamma}_k^{\text{cov}}(\psi_k)\lvert\psi_k\rangle=\int\!\! d\mu_k \langle\psi_k\rvert\bar{\Gamma}_k(\psi_k)\lvert\psi_k\rangle.\nonumber
\end{equation}
Therefore, it achieves the same value of $p$ (see Eq.~(\ref{eq:m1})) as $\bar{\Gamma}_k$. Thus, by convenience we consider that Alice and Bob implement the covariant map $\bar{\Gamma}_k^{\text{cov}}(\psi_k)$. In general, this is the depolarizing map \cite{NielsenandChuangbook}:
\begin{equation}
\bar{\Gamma}_k^{\text{cov}}(\phi)=\sum_{i=0}^{3}E_i\phi E_i^\dagger,\nonumber
\end{equation}
where $\phi\in\mathcal{D}\bigl(\mathbb{C}^2\bigr)$, $E_0=\lambda_k I$, $E_i=((1-\lambda_k)/3)\sigma_i$, $1/4\leq \lambda_k\leq 1$ and $\sigma_i$ are the Pauli matrices, for $i = 1, 2, 3$. Application of the depolarizing map to a qubit that is in the singlet state with another qubit, as in version I of the QIC game, gives as output the state $\omega_k$ given by Eq.~(\ref{eq:m7}) of the main text.

\subsection{A useful bound}
We show the bound
\begin{equation}
\label{eq:12}
\sum_{k=0}^{n-1}I\left(C_k:B_k\right)\leq I\left(C:B'\right),
\end{equation}
which will be useful to deduce an upper bound on $P$. The proof is equivalent to the one for classical bits \cite{ic}.

We notice that 
\begin{eqnarray}
\label{eq:13}
I\left(C:B'\right)&\equiv& I\left(C_0C_1\cdots C_{n-1}:B'\right)\nonumber\\
&=&I\left(C_0:B'\right)+I\left(C_1C_2\cdots C_{n-1}:B'C_0\right)\nonumber\\
&&-\: I\left(C_1C_2\cdots C_{n-1}:C_0\right).
\end{eqnarray}
Since Charlie's qubits are in a product state with each other, we have that
\begin{equation}
\label{eq:14}
I\left(C_1C_2\cdots C_{n-1}:C_0\right)=0.
\end{equation}
The data-processing inequality implies that
\begin{equation}
\label{eq:15}
I\left(C_1C_2\cdots C_{n-1}:B'C_0\right)\geq I\left(C_1C_2\cdots C_{n-1}:B'\right).
\end{equation}
From Eqs.~(\ref{eq:13})--(\ref{eq:15}) we obtain that
\begin{multline}
I\left(C_0C_1\cdots C_{n-1}:B'\right)\\
\geq I\left(C_0:B'\right)+I\left(C_1C_2\cdots C_{n-1}:B'\right).\nonumber
\end{multline}
After iterating these steps $n-1$ times, we have
\begin{equation}
\label{eq:16}
I\left(C:B'\right)\geq \sum_{k=0}^{n-1}I\left(C_k:B'\right).
\end{equation}
Since the system $B_k$ is output by Bob after local operations on his system $B'$, applying the data-processing inequality, we obtain $I(C_k:B')\geq I(C_k:B_k )$, which from Eq.~(\ref{eq:16}) implies Eq.~(\ref{eq:12}).

\subsection{Upper bound on $P$ from quantum information causality}
\begin{figure}[!h]
\includegraphics{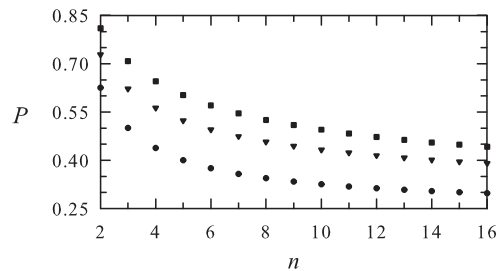}
 \caption{\label{fig2}Success probability ($P$) in the QIC game for ${m=1}$ achieved with the naive strategy, $P_{\text{N}}$ (circles), and with the best teleportation strategy that we have found, $P_{\text{T}}$ (triangles). The upper bound on $P$ obtained from quantum information causality, $P'$ (squares), is plotted too.}
\end{figure}

We show an upper bound on the success probability $P$ in the QIC game from quantum information causality:
\begin{equation}
\label{eq:17}
P\leq P',
\end{equation}
where we define $P'$ to be the maximum solution of the equation
\begin{equation}
\label{eq:18}
h(P')+(1-P')\log_23=2\left(1-\frac{m}{n}\right),
\end{equation}
and $h(x)=-x\log_2x-(1-x)\log_2(1-x)$ denotes the binary entropy. Some values of $P'$ are plotted in Fig.~\ref{fig2}.

We notice that since Charlie's and Bob's systems are initially uncorrelated, the quantum information causality bound (Eq.~(\ref{eq:m4}) of the main text) reduces to $I(C:B')\leq 2m$. Thus, from the bound given by Eq.~(\ref{eq:12}) we have that
\begin{equation}
\label{eq:19}
\sum_{k=0}^{n-1}I(C_k:B_k)\leq 2m.
\end{equation}

Charlie initially prepares the qubits $C_k$ and $A_k$ in the singlet state $\lvert\Psi^-\rangle_{C_kA_k }$, which after Alice's and Bob's operations is transformed into some state $\omega_k$, now in the joint system $C_kB_k$. We have shown that in general we can consider $\omega_k$ to be of the form given by Eq.~(\ref{eq:m7}) of the main text:
\begin{equation}
\omega_k=\lambda_k\Psi^-+\frac{1-\lambda_k}{3}\bigl(\Psi^++\Phi^++\Phi^-\bigr).\nonumber
\end{equation}
Thus, we have that $I(C_k:B_k)=2-S(\omega_k)$. Hence, from Eq.~(\ref{eq:19}) we have that
\begin{equation}
\label{eq:20}
\frac{1}{n}\sum_{k=0}^{n-1}S(\omega_k)\geq 2\left(1-\frac{m}{n}\right).
\end{equation}

We define the state $\omega\equiv\sum_{k=0}^{n-1}\omega_k/n$. From the concavity of the von Neumann entropy \cite{NielsenandChuangbook}, we obtain $S(\omega)\geq \sum_{k=0}^{n-1}S(\omega_k)/n$, which together with Eq.~(\ref{eq:20}) implies
\begin{equation}
\label{eq:21}
S(\omega)\geq 2\left(1-\frac{m}{n}\right).
\end{equation}
From the definitions of $P$ (Eq.~(\ref{eq:m2}) of the main text) and $\omega$, and the form of $\omega_k$ (Eq.~(\ref{eq:m7}) of the main text) we have that
\begin{equation}
\label{eq:22}
\omega=P\Psi^-+\frac{1-P}{3}\bigl(\Psi^++\Phi^++\Phi^-\bigr),
\end{equation}
which has von Neumann entropy $S(\omega)=h(P)+(1-P)\log_23$, where $h(x)=-x\log_2x-(1-x)\log_2(1-x)$ is the binary entropy. Thus, from Eq.~(\ref{eq:21}) we have that
\begin{equation}
\label{eq:23}
h(P)+(1-P)\log_23\geq 2\left(1-\frac{m}{n}\right),
\end{equation}
which implies Eq.~(\ref{eq:17}). This can be seen as follows. The function $h(P)+(1-P)\log_23$ corresponds to the Shannon entropy of a random variable taking four values, one with probability $P$ and the others with probability $(1-P)/3$ \cite{NielsenandChuangbook}. It is a strictly increasing function of $P$ in the range $[0,1/4]$ and a strictly decreasing function in the range $[1/4,1]$. It takes the values $\log_23$ at $P=0$ and $P=0.609$, 2 at $P=1/4$ and 0 at $P=1$. If $2(1-m/n)\geq \log_23$, Eq.~(\ref{eq:18}) has two solutions, one in the range $[0, 1/4]$ and the other one in the range $[1/4,0.609]$. Otherwise, Eq.~(\ref{eq:18}) has a single solution in the range $(0.609, 1]$. Therefore, the maximum solution of Eq.~(\ref{eq:18}) is in the range $[1/4, 1]$. Since in this range the function $h(P)+(1-P)\log_23$ is strictly decreasing, Eq.~(\ref{eq:23}) implies Eq.~(\ref{eq:17}).

In particular, we can easily see from Eq.~(\ref{eq:21}) that if $m< n$ then $S(\omega)>0$. Therefore, in this case $\omega$ cannot be a perfect singlet, which from Eq.~(\ref{eq:22}) implies that $P<1$.

\subsection{Upper bound on $Q$ for nonlocal strategies}
We have obtained an upper bound on the success probability $Q$ in the IC-2 game, defined in the main text, for a particular class of strategies in the case $m=1$:
\begin{equation}
\label{eq:24}
Q\leq Q',
\end{equation}
where $Q'\equiv\bigl(1 + 3n^{-1/2}\bigr)/4$. The considered class of strategies is the following. 

\emph{Nonlocal strategies in the IC-2 game}. Alice and Bob share an entangled state $\lvert\psi\rangle\in\mathcal{H}$. They perform a local projective measurement on their part of $\lvert\psi\rangle$. Alice chooses her measurement according to her value of $x\equiv(x_0,x_1,\ldots,x_{n-1})$. Recall that $x_j\equiv (x_j^0,x_j^1)$, for $j=0, 1, \ldots, n-1$. Bob chooses his measurement according to his number $k$. Their measurement outcomes are the two bit numbers $(a_k^0,a_k^1)$ and $(b_k^0,b_k^1)$, respectively. Alice sends Bob her outcome. Bob outputs the two bit value $y_k\equiv (y_k^0,y_k^1)$, where $y_k^j=a_k^j\oplus b_k^j$, for $j = 0, 1$, and $\oplus$ denotes sum modulo 2. The success probability is
\begin{equation}
Q=\frac{1}{n}\sum_{k=0}^{n-1}P\left(y_k^0=x_k^0,y_k^1=x_k^1\right).\nonumber
\end{equation}

This class of strategies is not general. For example, a more general strategy would be one in which Bob uses Alice's message in order to choose his measurement. 

It can easily be computed that for $m = 1$ and $n \geq 50$, $P' < Q',$ where $P'$ is defined by Eq.~(\ref{eq:18}). Therefore, the bound given by Eq.~(\ref{eq:24}) cannot be achieved for $n \geq 50$, otherwise Eq.~(\ref{eq:m5}) of the main text, and hence quantum information causality, could be violated by a teleportation strategy achieving $P = Q'$. 

Now we present the proof of Eq.~(\ref{eq:24}). This is an extension of the one given in Ref.~\cite{AS11} for the IC-1 game. Let $\mathcal{H}=\mathcal{H}_A\otimes\mathcal{H}_B$. Alice and Bob measure their respective systems, $A$ and $B$, in the orthonormal bases $\lbrace\lvert\nu_{r,s}^{x}\rangle\rbrace_{r,s=0}^1$ and $\lbrace\lvert w_{t,u}^{k}\rangle\rbrace_{t,u=0}^1$. After the measurement is completed, the state $\lvert\psi\rangle$ projects into the state $\lvert\nu_{a_k^0,a_k^1}^{x}\rangle\lvert w_{b_k^0,b_k^1}^{k}\rangle$. We define the Hermitian operators 
\begin{eqnarray}
\hat{A}_x&\equiv &\sum_{r=0}^1\sum_{s=0}^1(-1)^{r+s}\lvert\nu_{r,s}^x\rangle\langle\nu_{r,s}^x\rvert,\nonumber\\
\hat{B}_k&\equiv &\sum_{t=0}^1\sum_{u=0}^1(-1)^{t+u}\lvert w_{t,u}^k\rangle\langle w_{t,u}^k\rvert, \nonumber
\end{eqnarray}
acting on $\mathcal{H}_A$ and $\mathcal{H}_B$, respectively. We also define $E_{x,k}\equiv(-1)^{x_k^0+x_k^1}\langle\psi\rvert\hat{A}_x\hat{B}_k\lvert\psi\rangle$. Writing the state $\lvert\psi\rangle$ in the basis $\lbrace\lvert\nu_{r,s}^{x}\rangle\lvert w_{t,u}^{k}\rangle\rbrace_{r,s,t,u=0}^1$, using that $y_k^j=a_k^j\oplus b_k^j$, for $j=0,1$, and noticing that $x$ is a completely random variable of $4^n$ possible values, it is easy to obtain that
\begin{multline}
\label{eq:25}
\frac{1}{n}\sum_{k=0}^{n-1}\left[P\left(y_k^0=x_k^0,y_k^1=x_k^1\right)+P\left(y_k^0\neq x_k^0,y_k^1\neq x_k^1\right)\right]\\
=\frac{1}{2}\biggl(1+\frac{1}{n4^n}\sum_{x,k}E_{x,k}\biggr).
\end{multline}
Following the procedure of Ref.~\cite{AS11}, it is obtained that
\begin{equation}
\frac{1}{2}\biggl(1+\frac{1}{n4^n}\sum_{x,k}E_{x,k}\biggr)\leq\frac{1}{2}\biggl(1+\frac{1}{\sqrt{n}}\biggr),\nonumber
\end{equation}
which from Eq.~(\ref{eq:25}) implies
\begin{multline}
\label{eq:26}
\frac{1}{n}\sum_{k=0}^{n-1}\left[P\left(y_k^0=x_k^0,y_k^1=x_k^1\right)+P\left(y_k^0\neq x_k^0,y_k^1\neq x_k^1\right)\right]\\
\leq\frac{1}{2}\biggl(1+\frac{1}{\sqrt{n}}\biggr).
\end{multline}

Following a similar procedure, by defining $E_{x,k}^j\equiv(-1)^{x_k^j}\langle\psi\rvert\hat{A}_x^j\hat{B}_k^j\lvert\psi\rangle$, for $j = 0, 1$, in terms of the operators
\begin{eqnarray}
\hat{A}_x^0&\equiv&\sum_{r=0}^1\sum_{s=0}^1(-1)^{r}\lvert\nu_{r,s}^x\rangle\langle\nu_{r,s}^x\rvert,\nonumber\\
\hat{B}_k^0&\equiv&\sum_{t=0}^1\sum_{u=0}^1(-1)^{t}\lvert w_{t,u}^k\rangle\langle w_{t,u}^k\rvert,
\nonumber\\
\hat{A}_x^1&\equiv&\sum_{r=0}^1\sum_{s=0}^1(-1)^{s}\lvert\nu_{r,s}^x\rangle\langle\nu_{r,s}^x\rvert,\nonumber\\
\hat{B}_k^1&\equiv&\sum_{t=0}^1\sum_{u=0}^1(-1)^{u}\lvert w_{t,u}^k\rangle\langle w_{t,u}^k\rvert,\nonumber
\end{eqnarray}
it can be shown that
\begin{multline}
\label{eq:27}
\frac{1}{n}\sum_{k=0}^{n-1}\left[P\left(y_k^0=x_k^0,y_k^1=x_k^1\right)+P\left(y_k^0=x_k^0,y_k^1\neq x_k^1\right)\right]\\
\leq\frac{1}{2}\biggl(1+\frac{1}{\sqrt{n}}\biggr),
\end{multline}
and that
\begin{multline}
\label{eq:28}
\frac{1}{n}\sum_{k=0}^{n-1}\left[P\left(y_k^0=x_k^0,y_k^1=x_k^1\right)+P\left(y_k^0\neq x_k^0,y_k^1= x_k^1\right)\right]\\
\leq\frac{1}{2}\biggl(1+\frac{1}{\sqrt{n}}\biggr).
\end{multline}
Adding Eqs.~(\ref{eq:26})--(\ref{eq:28}), using normalization of probabilities and arranging terms we obtain that
\begin{equation}
\frac{1}{n}\sum_{k=0}^{n-1}P\left(y_k^0=x_k^0,y_k^1=x_k^1\right)\leq\frac{1}{4}\biggl(1+\frac{3}{\sqrt{n}}\biggr),\nonumber
\end{equation}
as claimed.

\begin{thebibliography}{40}%
\makeatletter
\providecommand \@ifxundefined [1]{%
 \@ifx{#1\undefined}
}%
\providecommand \@ifnum [1]{%
 \ifnum #1\expandafter \@firstoftwo
 \else \expandafter \@secondoftwo
 \fi
}%
\providecommand \@ifx [1]{%
 \ifx #1\expandafter \@firstoftwo
 \else \expandafter \@secondoftwo
 \fi
}%
\providecommand \natexlab [1]{#1}%
\providecommand \enquote  [1]{``#1''}%
\providecommand \bibnamefont  [1]{#1}%
\providecommand \bibfnamefont [1]{#1}%
\providecommand \citenamefont [1]{#1}%
\providecommand \href@noop [0]{\@secondoftwo}%
\providecommand \href [0]{\begingroup \@sanitize@url \@href}%
\providecommand \@href[1]{\@@startlink{#1}\@@href}%
\providecommand \@@href[1]{\endgroup#1\@@endlink}%
\providecommand \@sanitize@url [0]{\catcode `\\12\catcode `\$12\catcode
  `\&12\catcode `\#12\catcode `\^12\catcode `\_12\catcode `\%12\relax}%
\providecommand \@@startlink[1]{}%
\providecommand \@@endlink[0]{}%
\providecommand \url  [0]{\begingroup\@sanitize@url \@url }%
\providecommand \@url [1]{\endgroup\@href {#1}{\urlprefix }}%
\providecommand \urlprefix  [0]{URL }%
\providecommand \Eprint [0]{\href }%
\providecommand \doibase [0]{http://dx.doi.org/}%
\providecommand \selectlanguage [0]{\@gobble}%
\providecommand \bibinfo  [0]{\@secondoftwo}%
\providecommand \bibfield  [0]{\@secondoftwo}%
\providecommand \translation [1]{[#1]}%
\providecommand \BibitemOpen [0]{}%
\providecommand \bibitemStop [0]{}%
\providecommand \bibitemNoStop [0]{.\EOS\space}%
\providecommand \EOS [0]{\spacefactor3000\relax}%
\providecommand \BibitemShut  [1]{\csname bibitem#1\endcsname}%
\let\auto@bib@innerbib\@empty
\bibitem [{\citenamefont {Nielsen}\ and\ \citenamefont
  {Chuang}(2000)}]{NielsenandChuangbook}%
  \BibitemOpen
  \bibfield  {author} {\bibinfo {author} {\bibfnamefont {M.~A.}\ \bibnamefont
  {Nielsen}}\ and\ \bibinfo {author} {\bibfnamefont {I.~L.}\ \bibnamefont
  {Chuang}},\ }\href@noop {} {\emph {\bibinfo {title} {Quantum Computation and
  Quantum Information}}}\ (\bibinfo  {publisher} {Cambridge University Press},\
  \bibinfo {address} {Cambridge, UK},\ \bibinfo {year} {2000})\BibitemShut
  {NoStop}%
\bibitem [{\citenamefont {Wootters}\ and\ \citenamefont {Zurek}(1982)}]{WZ82}%
  \BibitemOpen
  \bibfield  {author} {\bibinfo {author} {\bibfnamefont {W.~K.}\ \bibnamefont
  {Wootters}}\ and\ \bibinfo {author} {\bibfnamefont {W.~H.}\ \bibnamefont
  {Zurek}},\ }\href
  {http://www.nature.com/nature/journal/v299/n5886/abs/299802a0.html}
  {\bibfield  {journal} {\bibinfo  {journal} {Nature (London)}\ }\textbf
  {\bibinfo {volume} {299}},\ \bibinfo {pages} {802} (\bibinfo {year}
  {1982})}\BibitemShut {NoStop}%
\bibitem [{\citenamefont {Dieks}(1982)}]{D82}%
  \BibitemOpen
  \bibfield  {author} {\bibinfo {author} {\bibfnamefont {D.}~\bibnamefont
  {Dieks}},\ }\href {http://dx.doi.org/10.1016/0375-9601(82)90084-6} {\bibfield
   {journal} {\bibinfo  {journal} {Phys. Lett. A}\ }\textbf {\bibinfo {volume}
  {92}},\ \bibinfo {pages} {271} (\bibinfo {year} {1982})}\BibitemShut
  {NoStop}%
\bibitem [{\citenamefont {Bennett}\ \emph {et~al.}(1993)\citenamefont
  {Bennett}, \citenamefont {Brassard}, \citenamefont {Cr\'epeau}, \citenamefont
  {Jozsa}, \citenamefont {Peres},\ and\ \citenamefont
  {Wootters}}]{teleportation}%
  \BibitemOpen
  \bibfield  {author} {\bibinfo {author} {\bibfnamefont {C.~H.}\ \bibnamefont
  {Bennett}}, \bibinfo {author} {\bibfnamefont {G.}~\bibnamefont {Brassard}},
  \bibinfo {author} {\bibfnamefont {C.}~\bibnamefont {Cr\'epeau}}, \bibinfo
  {author} {\bibfnamefont {R.}~\bibnamefont {Jozsa}}, \bibinfo {author}
  {\bibfnamefont {A.}~\bibnamefont {Peres}}, \ and\ \bibinfo {author}
  {\bibfnamefont {W.~K.}\ \bibnamefont {Wootters}},\ }\href
  {http://link.aps.org/doi/10.1103/PhysRevLett.70.1895} {\bibfield  {journal}
  {\bibinfo  {journal} {Phys. Rev. Lett.}\ }\textbf {\bibinfo {volume} {70}},\
  \bibinfo {pages} {1895} (\bibinfo {year} {1993})}\BibitemShut {NoStop}%
\bibitem [{\citenamefont {Bennett}\ and\ \citenamefont {Wiesner}(1992)}]{sdc}%
  \BibitemOpen
  \bibfield  {author} {\bibinfo {author} {\bibfnamefont {C.~H.}\ \bibnamefont
  {Bennett}}\ and\ \bibinfo {author} {\bibfnamefont {S.~J.}\ \bibnamefont
  {Wiesner}},\ }\href {http://link.aps.org/doi/10.1103/PhysRevLett.69.2881}
  {\bibfield  {journal} {\bibinfo  {journal} {Phys. Rev. Lett.}\ }\textbf
  {\bibinfo {volume} {69}},\ \bibinfo {pages} {2881} (\bibinfo {year}
  {1992})}\BibitemShut {NoStop}%
\bibitem [{\citenamefont {Bennett}\ and\ \citenamefont
  {Brassard}(1984)}]{BB84}%
  \BibitemOpen
  \bibfield  {author} {\bibinfo {author} {\bibfnamefont {C.~H.}\ \bibnamefont
  {Bennett}}\ and\ \bibinfo {author} {\bibfnamefont {G.}~\bibnamefont
  {Brassard}},\ }in\ \href@noop {} {\emph {\bibinfo {booktitle} {Proceedings of
  IEEE International Conference on Computers, Systems, and Signal Processing,
  Bangalore, India}}}\ (\bibinfo  {publisher} {IEEE},\ \bibinfo {address} {New
  York},\ \bibinfo {year} {1984})\ p.\ \bibinfo {pages} {175}\BibitemShut
  {NoStop}%
\bibitem [{\citenamefont {Ekert}(1991)}]{E91}%
  \BibitemOpen
  \bibfield  {author} {\bibinfo {author} {\bibfnamefont {A.~K.}\ \bibnamefont
  {Ekert}},\ }\href {http://link.aps.org/doi/10.1103/PhysRevLett.67.661}
  {\bibfield  {journal} {\bibinfo  {journal} {Phys. Rev. Lett.}\ }\textbf
  {\bibinfo {volume} {67}},\ \bibinfo {pages} {661} (\bibinfo {year}
  {1991})}\BibitemShut {NoStop}%
\bibitem [{\citenamefont {Barrett}\ \emph {et~al.}(2005)\citenamefont
  {Barrett}, \citenamefont {Hardy},\ and\ \citenamefont {Kent}}]{BHK05}%
  \BibitemOpen
  \bibfield  {author} {\bibinfo {author} {\bibfnamefont {J.}~\bibnamefont
  {Barrett}}, \bibinfo {author} {\bibfnamefont {L.}~\bibnamefont {Hardy}}, \
  and\ \bibinfo {author} {\bibfnamefont {A.}~\bibnamefont {Kent}},\ }\href
  {http://link.aps.org/doi/10.1103/PhysRevLett.95.010503} {\bibfield  {journal}
  {\bibinfo  {journal} {Phys. Rev. Lett.}\ }\textbf {\bibinfo {volume} {95}},\
  \bibinfo {pages} {010503} (\bibinfo {year} {2005})}\BibitemShut {NoStop}%
\bibitem [{\citenamefont {Einstein}\ \emph {et~al.}(1935)\citenamefont
  {Einstein}, \citenamefont {Podolsky},\ and\ \citenamefont {Rosen}}]{EPR35}%
  \BibitemOpen
  \bibfield  {author} {\bibinfo {author} {\bibfnamefont {A.}~\bibnamefont
  {Einstein}}, \bibinfo {author} {\bibfnamefont {B.}~\bibnamefont {Podolsky}},
  \ and\ \bibinfo {author} {\bibfnamefont {N.}~\bibnamefont {Rosen}},\ }\href
  {http://link.aps.org/doi/10.1103/PhysRev.47.777} {\bibfield  {journal}
  {\bibinfo  {journal} {Phys. Rev.}\ }\textbf {\bibinfo {volume} {47}},\
  \bibinfo {pages} {777} (\bibinfo {year} {1935})}\BibitemShut {NoStop}%
\bibitem [{\citenamefont {Bell}(1964)}]{Bell}%
  \BibitemOpen
  \bibfield  {author} {\bibinfo {author} {\bibfnamefont {J.~S.}\ \bibnamefont
  {Bell}},\ }\href@noop {} {\bibfield  {journal} {\bibinfo  {journal}
  {Physics}\ }\textbf {\bibinfo {volume} {1}},\ \bibinfo {pages} {195}
  (\bibinfo {year} {1964})}\BibitemShut {NoStop}%
\bibitem [{\citenamefont {Ghirardi}\ \emph {et~al.}(1980)\citenamefont
  {Ghirardi}, \citenamefont {Rimini},\ and\ \citenamefont {Weber}}]{GRW80}%
  \BibitemOpen
  \bibfield  {author} {\bibinfo {author} {\bibfnamefont {G.~C.}\ \bibnamefont
  {Ghirardi}}, \bibinfo {author} {\bibfnamefont {A.}~\bibnamefont {Rimini}}, \
  and\ \bibinfo {author} {\bibfnamefont {T.}~\bibnamefont {Weber}},\ }\href
  {http://www.springerlink.com/content/t5h287h1r26j25j4/} {\bibfield  {journal}
  {\bibinfo  {journal} {Lett. Nuov. Cim.}\ }\textbf {\bibinfo {volume} {27}},\
  \bibinfo {pages} {293} (\bibinfo {year} {1980})}\BibitemShut {NoStop}%
\bibitem [{\citenamefont {Kholevo}(1973)}]{K73}%
  \BibitemOpen
  \bibfield  {author} {\bibinfo {author} {\bibfnamefont {A.~S.}\ \bibnamefont
  {Kholevo}},\ }\href@noop {} {\bibfield  {journal} {\bibinfo  {journal}
  {Probl. Inf. Transm.}\ }\textbf {\bibinfo {volume} {9}},\ \bibinfo {pages}
  {177} (\bibinfo {year} {1973})}\BibitemShut {NoStop}%
\bibitem [{\citenamefont {Paw{\l}owski}\ \emph {et~al.}(2009)\citenamefont
  {Paw{\l}owski}, \citenamefont {Paterek}, \citenamefont {Kaszlikowski},
  \citenamefont {Scarani}, \citenamefont {Winter},\ and\ \citenamefont
  {{\.{Z}}ukowski}}]{ic}%
  \BibitemOpen
  \bibfield  {author} {\bibinfo {author} {\bibfnamefont {M.}~\bibnamefont
  {Paw{\l}owski}}, \bibinfo {author} {\bibfnamefont {T.}~\bibnamefont
  {Paterek}}, \bibinfo {author} {\bibfnamefont {D.}~\bibnamefont
  {Kaszlikowski}}, \bibinfo {author} {\bibfnamefont {V.}~\bibnamefont
  {Scarani}}, \bibinfo {author} {\bibfnamefont {A.}~\bibnamefont {Winter}}, \
  and\ \bibinfo {author} {\bibfnamefont {M.}~\bibnamefont {{\.{Z}}ukowski}},\
  }\href
  {http://www.nature.com/nature/journal/v461/n7267/full/nature08400.html}
  {\bibfield  {journal} {\bibinfo  {journal} {Nature (London)}\ }\textbf
  {\bibinfo {volume} {461}},\ \bibinfo {pages} {1101} (\bibinfo {year}
  {2009})}\BibitemShut {NoStop}%
\bibitem [{\citenamefont {Allcock}\ \emph {et~al.}(2009)\citenamefont
  {Allcock}, \citenamefont {Brunner}, \citenamefont {Paw{\l}owski},\ and\
  \citenamefont {Scarani}}]{ABPS09}%
  \BibitemOpen
  \bibfield  {author} {\bibinfo {author} {\bibfnamefont {J.}~\bibnamefont
  {Allcock}}, \bibinfo {author} {\bibfnamefont {N.}~\bibnamefont {Brunner}},
  \bibinfo {author} {\bibfnamefont {M.}~\bibnamefont {Paw{\l}owski}}, \ and\
  \bibinfo {author} {\bibfnamefont {V.}~\bibnamefont {Scarani}},\ }\href
  {http://link.aps.org/doi/10.1103/PhysRevA.80.040103} {\bibfield  {journal}
  {\bibinfo  {journal} {Phys. Rev. A}\ }\textbf {\bibinfo {volume} {80}},\
  \bibinfo {pages} {040103} (\bibinfo {year} {2009})}\BibitemShut {NoStop}%
\bibitem [{\citenamefont {Cavalcanti}\ \emph {et~al.}(2010)\citenamefont
  {Cavalcanti}, \citenamefont {Salles},\ and\ \citenamefont {Scarani}}]{CSS10}%
  \BibitemOpen
  \bibfield  {author} {\bibinfo {author} {\bibfnamefont {D.}~\bibnamefont
  {Cavalcanti}}, \bibinfo {author} {\bibfnamefont {A.}~\bibnamefont {Salles}},
  \ and\ \bibinfo {author} {\bibfnamefont {V.}~\bibnamefont {Scarani}},\ }\href
  {http://www.nature.com/ncomms/journal/v1/n9/full/ncomms1138.html} {\bibfield
  {journal} {\bibinfo  {journal} {Nat. Commun.}\ }\textbf {\bibinfo {volume}
  {1}},\ \bibinfo {pages} {136} (\bibinfo {year} {2010})}\BibitemShut {NoStop}%
\bibitem [{\citenamefont {Gallego}\ \emph {et~al.}(2011)\citenamefont
  {Gallego}, \citenamefont {W{\"{u}}rflinger}, \citenamefont {Ac\'in},\ and\
  \citenamefont {Navascu\'es}}]{GWAN11}%
  \BibitemOpen
  \bibfield  {author} {\bibinfo {author} {\bibfnamefont {R.}~\bibnamefont
  {Gallego}}, \bibinfo {author} {\bibfnamefont {L.~E.}\ \bibnamefont
  {W{\"{u}}rflinger}}, \bibinfo {author} {\bibfnamefont {A.}~\bibnamefont
  {Ac\'in}}, \ and\ \bibinfo {author} {\bibfnamefont {M.}~\bibnamefont
  {Navascu\'es}},\ }\href
  {http://link.aps.org/doi/10.1103/PhysRevLett.107.210403} {\bibfield
  {journal} {\bibinfo  {journal} {Phys. Rev. Lett.}\ }\textbf {\bibinfo
  {volume} {107}},\ \bibinfo {pages} {210403} (\bibinfo {year}
  {2011})}\BibitemShut {NoStop}%
\bibitem [{\citenamefont {Yang}\ \emph {et~al.}(2012)\citenamefont {Yang},
  \citenamefont {Cavalcanti}, \citenamefont {Almeida}, \citenamefont {Teo},\
  and\ \citenamefont {Scarani}}]{YCATS12}%
  \BibitemOpen
  \bibfield  {author} {\bibinfo {author} {\bibfnamefont {T.~H.}\ \bibnamefont
  {Yang}}, \bibinfo {author} {\bibfnamefont {D.}~\bibnamefont {Cavalcanti}},
  \bibinfo {author} {\bibfnamefont {M.~L.}\ \bibnamefont {Almeida}}, \bibinfo
  {author} {\bibfnamefont {C.}~\bibnamefont {Teo}}, \ and\ \bibinfo {author}
  {\bibfnamefont {V.}~\bibnamefont {Scarani}},\ }\href
  {http://iopscience.iop.org/1367-2630/14/1/013061} {\bibfield  {journal}
  {\bibinfo  {journal} {New J. Phys.}\ }\textbf {\bibinfo {volume} {14}},\
  \bibinfo {pages} {013061} (\bibinfo {year} {2012})}\BibitemShut {NoStop}%
\bibitem [{\citenamefont {Cirel'son}(1980)}]{C80}%
  \BibitemOpen
  \bibfield  {author} {\bibinfo {author} {\bibfnamefont {B.~S.}\ \bibnamefont
  {Cirel'son}},\ }\href {http://www.springerlink.com/content/l57053g573430450/}
  {\bibfield  {journal} {\bibinfo  {journal} {Lett. Math. Phys.}\ }\textbf
  {\bibinfo {volume} {4}},\ \bibinfo {pages} {93} (\bibinfo {year}
  {1980})}\BibitemShut {NoStop}%
\bibitem [{\citenamefont {Popescu}\ and\ \citenamefont
  {Rohrlich}(1994)}]{PR94}%
  \BibitemOpen
  \bibfield  {author} {\bibinfo {author} {\bibfnamefont {S.}~\bibnamefont
  {Popescu}}\ and\ \bibinfo {author} {\bibfnamefont {D.}~\bibnamefont
  {Rohrlich}},\ }\href {http://www.springerlink.com/content/j842v3324u512nx0/}
  {\bibfield  {journal} {\bibinfo  {journal} {Found. Phys.}\ }\textbf {\bibinfo
  {volume} {24}},\ \bibinfo {pages} {379} (\bibinfo {year} {1994})}\BibitemShut
  {NoStop}%
\bibitem [{Note1()}]{Note1}%
  \BibitemOpen
  \bibinfo {note} {A different question, investigated in Refs. \cite
  {CMMPPP12,SKB12} is how much entanglement can increase under local operations
  and quantum communication.}\BibitemShut {Stop}%
\bibitem [{\citenamefont {Henderson}\ and\ \citenamefont
  {Vedral}(2001)}]{HV01}%
  \BibitemOpen
  \bibfield  {author} {\bibinfo {author} {\bibfnamefont {L.}~\bibnamefont
  {Henderson}}\ and\ \bibinfo {author} {\bibfnamefont {V.}~\bibnamefont
  {Vedral}},\ }\href {http://iopscience.iop.org/0305-4470/34/35/315/}
  {\bibfield  {journal} {\bibinfo  {journal} {J. Phys. A: Math. Gen.}\ }\textbf
  {\bibinfo {volume} {34}},\ \bibinfo {pages} {6899} (\bibinfo {year}
  {2001})}\BibitemShut {NoStop}%
\bibitem [{\citenamefont {Ollivier}\ and\ \citenamefont {Zurek}(2001)}]{OZ02}%
  \BibitemOpen
  \bibfield  {author} {\bibinfo {author} {\bibfnamefont {H.}~\bibnamefont
  {Ollivier}}\ and\ \bibinfo {author} {\bibfnamefont {W.~H.}\ \bibnamefont
  {Zurek}},\ }\href {http://link.aps.org/doi/10.1103/PhysRevLett.88.017901}
  {\bibfield  {journal} {\bibinfo  {journal} {Phys. Rev. Lett.}\ }\textbf
  {\bibinfo {volume} {88}},\ \bibinfo {pages} {017901} (\bibinfo {year}
  {2001})}\BibitemShut {NoStop}%
\bibitem [{\citenamefont {Groisman}\ \emph {et~al.}(2005)\citenamefont
  {Groisman}, \citenamefont {Popescu},\ and\ \citenamefont {Winter}}]{GPW05}%
  \BibitemOpen
  \bibfield  {author} {\bibinfo {author} {\bibfnamefont {B.}~\bibnamefont
  {Groisman}}, \bibinfo {author} {\bibfnamefont {S.}~\bibnamefont {Popescu}}, \
  and\ \bibinfo {author} {\bibfnamefont {A.}~\bibnamefont {Winter}},\ }\href
  {http://link.aps.org/doi/10.1103/PhysRevA.72.032317} {\bibfield  {journal}
  {\bibinfo  {journal} {Phys. Rev. A}\ }\textbf {\bibinfo {volume} {72}},\
  \bibinfo {pages} {032317} (\bibinfo {year} {2005})}\BibitemShut {NoStop}%
\bibitem [{Note2()}]{Note2}%
  \BibitemOpen
  \bibinfo {note} {Note that Refs.~\cite {HV01,OZ02,GPW05} propose measures for
  the purely classical and purely quantum parts of the correlations between two
  quantum systems, whose sum is equal to the quantum mutual information (see
  Ref. \cite {MBCPV12} for a review). We do not consider such a classification
  in our discussion.}\BibitemShut {Stop}%
\bibitem [{\citenamefont {Lanford}\ and\ \citenamefont
  {Robinson}(1968)}]{LR68}%
  \BibitemOpen
  \bibfield  {author} {\bibinfo {author} {\bibfnamefont {O.~E.}\ \bibnamefont
  {Lanford}}\ and\ \bibinfo {author} {\bibfnamefont {D.~W.}\ \bibnamefont
  {Robinson}},\ }\href {http://jmp.aip.org/resource/1/jmapaq/v9/i7/p1120_s1}
  {\bibfield  {journal} {\bibinfo  {journal} {J. Math. Phys.}\ }\textbf
  {\bibinfo {volume} {9}},\ \bibinfo {pages} {1120} (\bibinfo {year}
  {1968})}\BibitemShut {NoStop}%
\bibitem [{\citenamefont {Araki}\ and\ \citenamefont {Lieb}(1970)}]{AL70}%
  \BibitemOpen
  \bibfield  {author} {\bibinfo {author} {\bibfnamefont {H.}~\bibnamefont
  {Araki}}\ and\ \bibinfo {author} {\bibfnamefont {E.~H.}\ \bibnamefont
  {Lieb}},\ }\href {http://www.springerlink.com/content/wk224145h1441527/}
  {\bibfield  {journal} {\bibinfo  {journal} {Commun. Math. Phys.}\ }\textbf
  {\bibinfo {volume} {18}},\ \bibinfo {pages} {160} (\bibinfo {year}
  {1970})}\BibitemShut {NoStop}%
\bibitem [{\citenamefont {Hardy}()}]{H01}%
  \BibitemOpen
  \bibfield  {author} {\bibinfo {author} {\bibfnamefont {L.}~\bibnamefont
  {Hardy}},\ }\href@noop {} {}\Eprint {http://arxiv.org/abs/quant-ph/0101012}
  {arXiv:quant-ph/0101012} \BibitemShut {NoStop}%
\bibitem [{\citenamefont {Barrett}(2007)}]{B07}%
  \BibitemOpen
  \bibfield  {author} {\bibinfo {author} {\bibfnamefont {J.}~\bibnamefont
  {Barrett}},\ }\href {http://link.aps.org/doi/10.1103/PhysRevA.75.032304}
  {\bibfield  {journal} {\bibinfo  {journal} {Phys. Rev. A}\ }\textbf {\bibinfo
  {volume} {75}},\ \bibinfo {pages} {032304} (\bibinfo {year}
  {2007})}\BibitemShut {NoStop}%
\bibitem [{\citenamefont {Barnum}\ \emph {et~al.}(2007)\citenamefont {Barnum},
  \citenamefont {Barrett}, \citenamefont {Leifer},\ and\ \citenamefont
  {Wilce}}]{BBLW07}%
  \BibitemOpen
  \bibfield  {author} {\bibinfo {author} {\bibfnamefont {H.}~\bibnamefont
  {Barnum}}, \bibinfo {author} {\bibfnamefont {J.}~\bibnamefont {Barrett}},
  \bibinfo {author} {\bibfnamefont {M.}~\bibnamefont {Leifer}}, \ and\ \bibinfo
  {author} {\bibfnamefont {A.}~\bibnamefont {Wilce}},\ }\href
  {http://link.aps.org/doi/10.1103/PhysRevLett.99.240501} {\bibfield  {journal}
  {\bibinfo  {journal} {Phys. Rev. Lett.}\ }\textbf {\bibinfo {volume} {99}},\
  \bibinfo {pages} {240501} (\bibinfo {year} {2007})}\BibitemShut {NoStop}%
\bibitem [{\citenamefont {Short}\ and\ \citenamefont {Wehner}(2010)}]{SW10}%
  \BibitemOpen
  \bibfield  {author} {\bibinfo {author} {\bibfnamefont {A.~J.}\ \bibnamefont
  {Short}}\ and\ \bibinfo {author} {\bibfnamefont {S.}~\bibnamefont {Wehner}},\
  }\href {http://iopscience.iop.org/1367-2630/12/3/033023} {\bibfield
  {journal} {\bibinfo  {journal} {New J. Phys.}\ }\textbf {\bibinfo {volume}
  {12}},\ \bibinfo {pages} {033023} (\bibinfo {year} {2010})}\BibitemShut
  {NoStop}%
\bibitem [{\citenamefont {Barnum}\ \emph {et~al.}(2010)\citenamefont {Barnum},
  \citenamefont {Barrett}, \citenamefont {Clark}, \citenamefont {Leifer},
  \citenamefont {Spekkens}, \citenamefont {Stepanik}, \citenamefont {Wilce},\
  and\ \citenamefont {Wilke}}]{BBCLSSWW10}%
  \BibitemOpen
  \bibfield  {author} {\bibinfo {author} {\bibfnamefont {H.}~\bibnamefont
  {Barnum}}, \bibinfo {author} {\bibfnamefont {J.}~\bibnamefont {Barrett}},
  \bibinfo {author} {\bibfnamefont {L.~O.}\ \bibnamefont {Clark}}, \bibinfo
  {author} {\bibfnamefont {M.}~\bibnamefont {Leifer}}, \bibinfo {author}
  {\bibfnamefont {R.}~\bibnamefont {Spekkens}}, \bibinfo {author}
  {\bibfnamefont {N.}~\bibnamefont {Stepanik}}, \bibinfo {author}
  {\bibfnamefont {A.}~\bibnamefont {Wilce}}, \ and\ \bibinfo {author}
  {\bibfnamefont {R.}~\bibnamefont {Wilke}},\ }\href
  {http://iopscience.iop.org/1367-2630/12/3/033024} {\bibfield  {journal}
  {\bibinfo  {journal} {New J. Phys.}\ }\textbf {\bibinfo {volume} {12}},\
  \bibinfo {pages} {033024} (\bibinfo {year} {2010})}\BibitemShut {NoStop}%
\bibitem [{\citenamefont {Al-Safi}\ and\ \citenamefont {Short}(2011)}]{AS11}%
  \BibitemOpen
  \bibfield  {author} {\bibinfo {author} {\bibfnamefont {S.~W.}\ \bibnamefont
  {Al-Safi}}\ and\ \bibinfo {author} {\bibfnamefont {A.~J.}\ \bibnamefont
  {Short}},\ }\href {http://link.aps.org/doi/10.1103/PhysRevA.84.042323}
  {\bibfield  {journal} {\bibinfo  {journal} {Phys. Rev. A}\ }\textbf {\bibinfo
  {volume} {84}},\ \bibinfo {pages} {042323} (\bibinfo {year}
  {2011})}\BibitemShut {NoStop}%
\bibitem [{\citenamefont {Dahlsten}\ \emph {et~al.}(2012)\citenamefont
  {Dahlsten}, \citenamefont {Lercher},\ and\ \citenamefont {Renner}}]{DLR12}%
  \BibitemOpen
  \bibfield  {author} {\bibinfo {author} {\bibfnamefont {O.~C.~O.}\
  \bibnamefont {Dahlsten}}, \bibinfo {author} {\bibfnamefont {D.}~\bibnamefont
  {Lercher}}, \ and\ \bibinfo {author} {\bibfnamefont {R.}~\bibnamefont
  {Renner}},\ }\href {http://iopscience.iop.org/1367-2630/14/6/063024/}
  {\bibfield  {journal} {\bibinfo  {journal} {New J. Phys.}\ }\textbf {\bibinfo
  {volume} {14}},\ \bibinfo {pages} {063024} (\bibinfo {year}
  {2012})}\BibitemShut {NoStop}%
\bibitem [{\citenamefont {Masanes}\ \emph {et~al.}()\citenamefont {Masanes},
  \citenamefont {Mueller}, \citenamefont {Augusiak},\ and\ \citenamefont
  {Perez-Garcia}}]{MMAP12}%
  \BibitemOpen
  \bibfield  {author} {\bibinfo {author} {\bibfnamefont {L.}~\bibnamefont
  {Masanes}}, \bibinfo {author} {\bibfnamefont {M.~P.}\ \bibnamefont
  {Mueller}}, \bibinfo {author} {\bibfnamefont {R.}~\bibnamefont {Augusiak}}, \
  and\ \bibinfo {author} {\bibfnamefont {D.}~\bibnamefont {Perez-Garcia}},\
  }\href@noop {} {}\Eprint {http://arxiv.org/abs/1208.0493} {arXiv:1208.0493}
  \BibitemShut {NoStop}%
\bibitem [{\citenamefont {Wiesner}(1983)}]{W83}%
  \BibitemOpen
  \bibfield  {author} {\bibinfo {author} {\bibfnamefont {S.}~\bibnamefont
  {Wiesner}},\ }\href {http://dl.acm.org/citation.cfm?doid=1008908.1008920}
  {\bibfield  {journal} {\bibinfo  {journal} {SIGACT News}\ }\textbf {\bibinfo
  {volume} {15}},\ \bibinfo {pages} {78} (\bibinfo {year} {1983})}\BibitemShut
  {NoStop}%
\bibitem [{\citenamefont {Ambainis}\ \emph {et~al.}(2002)\citenamefont
  {Ambainis}, \citenamefont {Nayak}, \citenamefont {Ta-Shma},\ and\
  \citenamefont {Vazirani}}]{ANTV02}%
  \BibitemOpen
  \bibfield  {author} {\bibinfo {author} {\bibfnamefont {A.}~\bibnamefont
  {Ambainis}}, \bibinfo {author} {\bibfnamefont {A.}~\bibnamefont {Nayak}},
  \bibinfo {author} {\bibfnamefont {A.}~\bibnamefont {Ta-Shma}}, \ and\
  \bibinfo {author} {\bibfnamefont {U.}~\bibnamefont {Vazirani}},\ }\href
  {http://dl.acm.org/citation.cfm?doid=581771.581773} {\bibfield  {journal}
  {\bibinfo  {journal} {J. ACM}\ }\textbf {\bibinfo {volume} {49}},\ \bibinfo
  {pages} {496} (\bibinfo {year} {2002})}\BibitemShut {NoStop}%
\bibitem [{\citenamefont {Paw{\l}owski}\ and\ \citenamefont
  {{\.{Z}}ukowski}(2010)}]{PZ10}%
  \BibitemOpen
  \bibfield  {author} {\bibinfo {author} {\bibfnamefont {M.}~\bibnamefont
  {Paw{\l}owski}}\ and\ \bibinfo {author} {\bibfnamefont {M.}~\bibnamefont
  {{\.{Z}}ukowski}},\ }\href
  {http://link.aps.org/doi/10.1103/PhysRevA.81.042326} {\bibfield  {journal}
  {\bibinfo  {journal} {Phys. Rev. A}\ }\textbf {\bibinfo {volume} {81}},\
  \bibinfo {pages} {042326} (\bibinfo {year} {2010})}\BibitemShut {NoStop}%
\bibitem [{\citenamefont {Chuan}\ \emph {et~al.}(2012)\citenamefont {Chuan},
  \citenamefont {Maillard}, \citenamefont {Modi}, \citenamefont {Paterek},
  \citenamefont {Paternostro},\ and\ \citenamefont {Piani}}]{CMMPPP12}%
  \BibitemOpen
  \bibfield  {author} {\bibinfo {author} {\bibfnamefont {T.~K.}\ \bibnamefont
  {Chuan}}, \bibinfo {author} {\bibfnamefont {J.}~\bibnamefont {Maillard}},
  \bibinfo {author} {\bibfnamefont {K.}~\bibnamefont {Modi}}, \bibinfo {author}
  {\bibfnamefont {T.}~\bibnamefont {Paterek}}, \bibinfo {author} {\bibfnamefont
  {M.}~\bibnamefont {Paternostro}}, \ and\ \bibinfo {author} {\bibfnamefont
  {M.}~\bibnamefont {Piani}},\ }\href {\doibase 10.1103/PhysRevLett.109.070501}
  {\bibfield  {journal} {\bibinfo  {journal} {Phys. Rev. Lett.}\ }\textbf
  {\bibinfo {volume} {109}},\ \bibinfo {pages} {070501} (\bibinfo {year}
  {2012})}\BibitemShut {NoStop}%
\bibitem [{\citenamefont {Streltsov}\ \emph {et~al.}(2012)\citenamefont
  {Streltsov}, \citenamefont {Kampermann},\ and\ \citenamefont
  {Bru\ss{}}}]{SKB12}%
  \BibitemOpen
  \bibfield  {author} {\bibinfo {author} {\bibfnamefont {A.}~\bibnamefont
  {Streltsov}}, \bibinfo {author} {\bibfnamefont {H.}~\bibnamefont
  {Kampermann}}, \ and\ \bibinfo {author} {\bibfnamefont {D.}~\bibnamefont
  {Bru\ss{}}},\ }\href {\doibase 10.1103/PhysRevLett.108.250501} {\bibfield
  {journal} {\bibinfo  {journal} {Phys. Rev. Lett.}\ }\textbf {\bibinfo
  {volume} {108}},\ \bibinfo {pages} {250501} (\bibinfo {year}
  {2012})}\BibitemShut {NoStop}%
\bibitem [{\citenamefont {Modi}\ \emph {et~al.}(2012)\citenamefont {Modi},
  \citenamefont {Brodutch}, \citenamefont {Cable}, \citenamefont {Paterek},\
  and\ \citenamefont {Vedral}}]{MBCPV12}%
  \BibitemOpen
  \bibfield  {author} {\bibinfo {author} {\bibfnamefont {K.}~\bibnamefont
  {Modi}}, \bibinfo {author} {\bibfnamefont {A.}~\bibnamefont {Brodutch}},
  \bibinfo {author} {\bibfnamefont {H.}~\bibnamefont {Cable}}, \bibinfo
  {author} {\bibfnamefont {T.}~\bibnamefont {Paterek}}, \ and\ \bibinfo
  {author} {\bibfnamefont {V.}~\bibnamefont {Vedral}},\ }\href {\doibase
  10.1103/RevModPhys.84.1655} {\bibfield  {journal} {\bibinfo  {journal} {Rev.
  Mod. Phys.}\ }\textbf {\bibinfo {volume} {84}},\ \bibinfo {pages} {1655}
  (\bibinfo {year} {2012})}\BibitemShut {NoStop}%
\end{thebibliography}
\end{document}